\documentclass[showpacs,preprintnumbers,amsmath,amssymb,twocolumn,aps,10pt,prl]{revtex4-1}
\usepackage{default}
\usepackage{amssymb}
\usepackage{color}
\usepackage{amsmath} 
\usepackage{mathrsfs}
\usepackage{graphicx}
\usepackage{epsfig}

\begin{document}
\title{Front interaction induces excitable behavior}

\author{P. Parra-Rivas$^{1,2}$, M. A. Mat\'{\i}as$^{1}$, P. Colet$^{1}$, L. Gelens$^{2,3}$, D. Walgraef$^{1}$ and D. Gomila$^{1}$}

\affiliation{
$^{1}$Instituto de F\'{\i}sica Interdisciplinar y  Sistemas Complejos, IFISC (CSIC-UIB), Campus Universitat de les Illes
  Balears, E-07122 Palma de Mallorca, Spain\\
$^2$Applied Physics Research Group, APHY, Vrije Universiteit Brussel, 1050 Brussels, Belgium\\
$^3$Laboratory of Dynamics in Biological Systems, Department of Cellular and Molecular Medicine, KU Leuven, University of Leuven, B-3000 Leuven, Belgium
}
\date{\today}


\begin{abstract}
Spatially extended systems can support local transient excitations in which just a part of the system is excited. 
The mechanisms reported so far are local excitability and excitation of a localized structure. 
Here we introduce an alternative mechanism based on 
the coexistence of two homogeneous stable states and spatial coupling. We show the existence of a threshold for perturbations of the homogeneous state. Sub-threshold perturbations decay exponentially. Super-threshold perturbations induce the emergence of a long-lived structure formed by two back to back fronts that join the two homogeneous states. While in typical excitability the trajectory follows the remnants of a limit cycle, here reinjection is provided by front interaction, such that fronts slowly approach each other until eventually annihilating. This front-mediated mechanism shows that extended systems with no oscillatory regimes can display 
excitability. 
\end{abstract}
\maketitle

Excitability is a concept that originally comes from biology, inspired by the behavior of neurons and heart cells  \cite{Winfree_book}.
An excitable system is characterized by exhibiting a stable steady behavior, while responding to perturbations (e.g.
external stimuli) in two different ways: for stimuli below a certain threshold the system decays exponentially to the steady state, while for stimuli above threshold it exhibits a nontrivial excursion in phase space before decaying back to the steady state.
Interestingly, the 
excitable properties of neurons confer them computational capabilities, allowing 
them to process external inputs \cite{Kochbook}.

Excitable behavior has been classified in two types \cite{Izhikevich2000,Izhikevichbook}. Type I excitability arises when, changing a parameter, a limit cycle is destroyed by suitable global bifurcations (homoclinic or SNIC (Saddle-Node on the Invariant Circle)). Type II arises when a limit cycle is destroyed by a supercritical Hopf bifurcation preceded by a canard or by a fold of cycles (usually preceded by a subcritical Hopf). In both types, excitability can only appear after a limit cycle of non-zero amplitude is destroyed when changing a parameter. The excitable trajectory would follow the remnants of the cycle and ultimately 
return to the steady state  \cite{Izhikevich2000,Izhikevichbook}.

Several kinds of spatially extended systems display excitable behavior.
In the most straightforward case, systems which are individually excitable are coupled in space. These excitable media typically exhibit characteristic excitable waves or pulses \cite{Murraybook,Mikhailov_book,Meron1992}. Excitable 
behavior can also arise in a more subtle way, however, through the emergent dynamics of coherent 
structures, which does not require local excitable dynamics \cite{Gomila2005}. The excitable excursion follows the remnants of a limit cycle corresponding to an oscillatory localized structure and can be described in terms of an effective reduced phase space \cite{Gomila2005,Gomila2007,*Jacobo2008,*Parra2013,*Leo2013}. 

In particular in the context of cellular biology, transient localized excitations, also called patches, have been observed in early stages of cell migration. One of the most studied examples is the cellular slime mold, {\it Dyctiostellium discoideum}. In this system, the uniform application of the chemoattractant cAMP leads to the spontaneous emergence of localized regions of
high protein concentration, patches, that after some time dismantle and appear elsewhere \cite{Postma2003}. In
\cite{Hecht2010} an explanation was suggested in terms of a spatially extended model with local FitzHugh-Nagumo dynamics in the excitable regime. However, as pointed out in \cite{Huang2013}, 
in cell migration, direct evidence for excitability is lacking.

In this Letter we present an alternative mechanism leading to transient patches that requires neither local excitability, nor 
oscillatory localized structures, not even the existence of the remnants of a limit cycle, 
widening the classes of extended systems 
that can present excitability. In particular we show that only two simple 
ingredients are necessary: bistability between two homogeneous stable steady states (HSSSs) and spatial coupling allowing for monotonic fronts connecting these two states.

To illustrate the excitability mechanism introduced here we consider a prototypical model 
displaying bistability, namely, the Ginzburg-Landau equation for a real field $u\equiv u(x,t)$ in one spatial dimension $x$
\begin{equation}
 \partial_t u=\mu u-u^3+\partial_x^2 u .
 \label{GL}
 \end{equation}
The system is variational and therefore does not have oscillatory solutions. For $\mu>0$ the system has two equivalent HSSSs $u_{\pm}=\pm\sqrt{\mu}$ and stable fronts that connect them. The trivial solution $u=0$ is unstable and plays the role of a separatrix for the local dynamics. Here we take $\mu=1$ and thus $u_\pm= \pm 1$. There are two possible front solutions, known as kink and antikink, with opposite polarity and a monotonic $\tanh$ shape \cite{Coullet1987}. Fronts with opposite polarity attract each other with a strength that decays exponentially with the front separation \cite{Coullet1987}, and they ultimately annihilate in a behavior known as coarsening \cite{coarsening}. 

The rationale of the excitability mechanism is as follows.
While the system is sitting on a HSSS, small localized perturbations decay exponentially. Instead, for perturbations exceeding $u(x)=0$ in a wide enough spatial region, part of the system will initially evolve to the other (attracting) HSSS leading to the formation of a pair of kink-antikink fronts. In a second stage the two fronts interact, slowly approaching each other. If the resulting kink-antikink structure is relatively broad this second stage will be long-lived. Finally, in a third stage, kink and antikink annihilate each other and the system returns to the initial HSSS. These structures can be viewed as excitable excursions and, following \cite{Hecht2010}, we will refer to them as ``patches''

\begin{figure}[t!]
\includegraphics[width=\columnwidth]{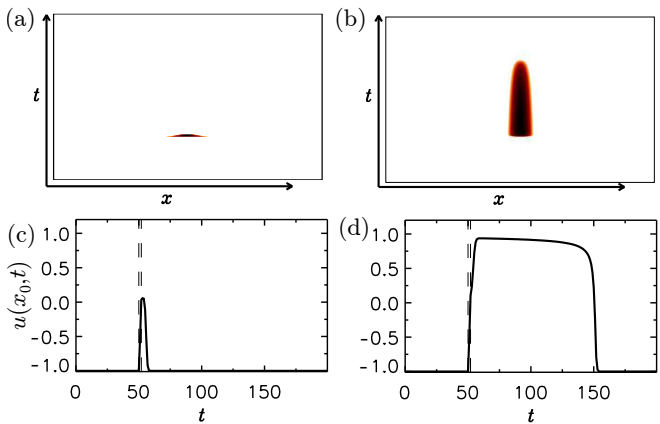}
\caption{(Color online) Evolution of $u(x,t)$ after a perturbation on $u_{-}$ with $\Gamma=15$, $t_0=50$ and $\Delta t=2$.
(a) and (c) correspond to a sub-threshold perturbation with $G=0.79$. (b) and (d) correspond to a super-threshold perturbation with 
$G=0.81$. The vertical dashed lines indicate the switching on and off of the perturbation. In panels (a) and (b) the range of $x$ is $[-22,22]$ and the range of $t$ is $ [0,200]$.}
\label{exc}
\end{figure}

The excitable behavior is illustrated in Fig.~\ref{exc} showing the spatio-temporal dynamics after perturbing $u_{-}$ during a time $\Delta t$ with a 
Gaussian spatial profile that is added to Eq.~(\ref{GL}):
\begin{equation} 
 g(x)=  G\exp\left[-(\ln2) (x-x_{0})^2/\Gamma^{2}\right],
 \label{prof}
\end{equation}
where $G$ and $x_0$ are the height and center of the Gaussian and $\Gamma$ is the half width at half maximum. For low enough values of $G$ the perturbation decays exponentially [panels (a) and (c)]. In contrast, for a slightly larger $G$ [panels (b) and (d)], $u(x_0,t)$ closely approaches $u_{+}$ and a long-lived patch is formed before returning to $u_-$. The existence of a threshold in the perturbation size separating two different ways of responding is a clear indication of excitability. 

\begin{figure}[t!]
\includegraphics[width=\columnwidth]{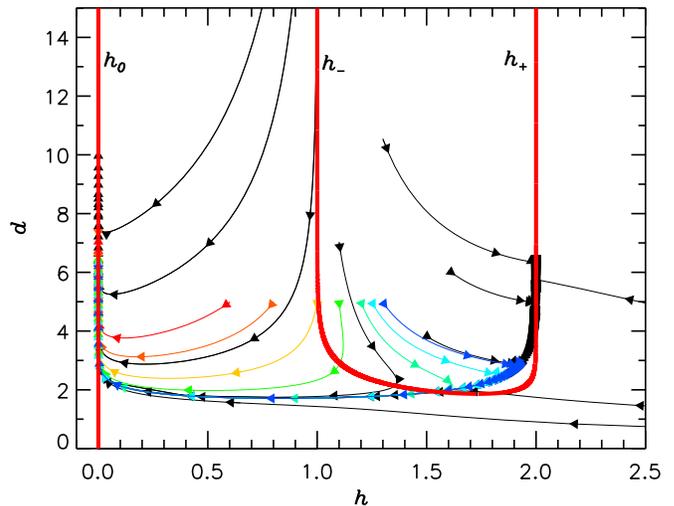}
\caption{(Color online) Projection of the dynamics into the $(h,d)$ phase space. Red lines correspond to the analytical $h$-nullclines (\ref{walgraef}). 
Lines with arrows show trajectories obtained integrating Eq.~(\ref{GL}) starting from an initial condition (\ref{eq:initial_condition}) with different $d_0$ and $H_0$. The arrows correspond to the velocity field, allowing to identify fast and slow time scales. Trajectories in color correspond to $d_0=4.98$ and different $H_0$, whose temporal evolution is plotted in Fig.~\ref{time}.} 
\label{PP_d_H}
\end{figure}

In {\it classical\/} Type II excitable 2-D systems (e.g. 
the FitzHugh-Nagumo model), the excitable behavior can be understood \cite{Gerstner} by analyzing the shape of 
the nullclines in phase space \footnote{Nullclines are defined as the 
geometric place in which the time derivative of one of the system variables is 
zero, being found the fixed points at the intersection of these nullclines.}.
Here, despite the fact that the phase space is infinite dimensional, 
we find a very similar scenario by considering a 2-D effective phase space $(h,d)$ 
where $h(t)\equiv u(x_0,t)-u_-$ is the height of the patch at its center $x_0$, and $d(t)$ its half width at half maximum.
For $d$ larger than the front width, we can approximate the shape of the patch by two $\tanh$ fronts placed back to back
\begin{equation}
 u(x,t)=u_-+h(t)\left[\eta(x,x_0-d(t))-\eta(x,x_0+d(t))\right]
 \label{eq:kink-antikink}
\end{equation}
where 
$\eta(x,a)\equiv\tanh(({x-a})/{\sqrt{2}} )$. 

Introducing (\ref{eq:kink-antikink}) in Eq.~(\ref{GL}), evaluating it at $x=x_0$, we get 
\begin{equation}
 \dot{h}=h\left[6h\tanh\left(\frac{d}{\sqrt{2}}\right)+(1-4h^2)\tanh^2\left(\frac{d}{\sqrt{2}}\right)-3\right],
 \label{eq:h}
\end{equation}
where we have considered that the evolution of $d(t)$ is much slower than that of $h(t)$. The $h$-nullcline is 
\begin{equation}
h_{\pm}=\frac{3}{4} \coth(d/\sqrt{2}) \pm \frac{1}{4} \sqrt{4-3\coth^2(d/\sqrt{2})},\; \; h_0=0.
 \label{walgraef}
\end{equation}
It is composed of three pieces, $h_+$ which is attracting and has a vertical asymptote at $h=2$, $h_-$ repelling with a vertical asymptote at $h=1$, and $h_0$ also stable. 
$h_+$ and $h_-$ are connected at $h=\sqrt{3}/2$, $d=\ln(7+4\sqrt{3})/\sqrt{2}\approx1.86$. $h$-nullclines are shown in Fig.~\ref{PP_d_H}, which also displays the projection of the evolution obtained from numerical integration of Eq.~(\ref{GL}) after perturbing the $u_-$ HSSS. Here, instead of adding a signal for a short time to induce perturbations, we now perturb the initial spatial profile:  
\begin{equation}
 u(x,0)=u_-+ H_0\exp\left[-(\ln2) (x-x_{0})^2/d_0^{2}\right] .
 \label{eq:initial_condition}
\end{equation}

For $h>0$, the derivative of $h$ is negative, except inside the U-shaped region delimited by $h_-$ and $h_+$ where $\dot{h}>0$. 
For initial conditions located to the right of $h_-$ the trajectory evolves rapidly towards the nullcline $h_+$. This corresponds to the first stage of the excitable excursion. The center of the patch is very flat and evolves towards $u_+$ on a time scale of order O($\mu$) making $h$ the fast variable of the dynamics, while $d$ changes at a much smaller rate. The outcome of this stage is the formation of a kink-antikink pair connecting $u_-$ with $u_+$ and then again back to $u_-$. 
In the reduced phase space this means that the nullcline $h_+$ has been reached. After reaching the nullcline, the patch evolves slowly along $h_+$. This corresponds to the second stage of the excitable excursion in which kink and antikink slowly approach each other decreasing $d$ following \cite{Coullet1987}
\begin{equation}
 \dot{d}=c\exp(-\gamma d) ,
 \label{eq:expd}
\end{equation}
where $c=-24\sqrt{2\mu}$ and $\gamma=\sqrt{2\mu}$ \cite{Gelens2010}. 
Finally the third stage in which the kink and antikink annihilate each other corresponds to the fast jump to nullcline $h_0$. This reinjection mechanism is not following the remnants of a limit cycle in phase space since the system Eq.~(\ref{GL}) does not have periodic solutions. For initial conditions located at the left of $h_-$ or below the connection of $h_-$ with $h_+$, the system evolves quickly to $h_0$. This corresponds to sub-threshold perturbations which decay exponentially. 

\begin{figure}[t!]
\includegraphics[width=\columnwidth]{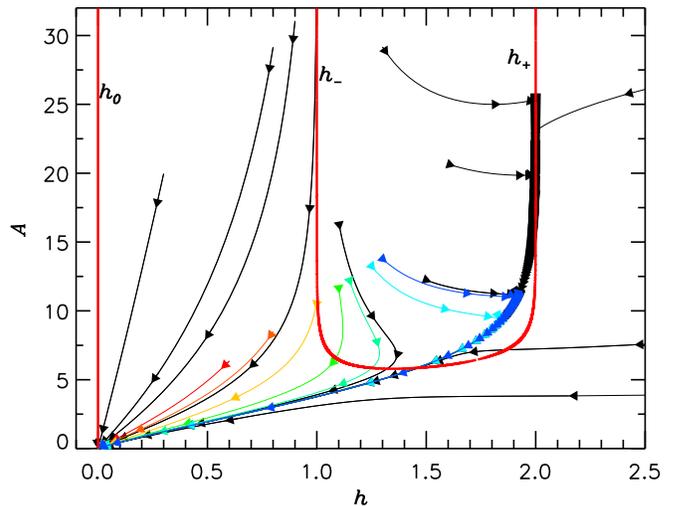}
\caption{(Color online) Phase portrait as Fig.~\ref{PP_d_H} in the $(h,A)$ plane.} 
\label{PP_A_H}
\end{figure}

A drawback of the $(h,d)$ description is that the width of the patch is not well defined for $h=0$. As a consequence it is not evident in Fig.~\ref{PP_d_H} that all the trajectories finally evolve to $u_-$. A convenient way to avoid this is to use the area of the patch $A$ instead of $d$ as shown in Fig.~\ref{PP_A_H}). To estimate the nullclines theoretically we have used $A=4hd[1-2\exp(-L/\sqrt{2})]$, being $L$ the size of the system. In this representation all trajectories ultimately converge to the fixed point $(h,A)=(0,0)$ which corresponds to $u_-$. 

Altogether, our analysis shows that the $h$-nullclines given by (\ref{walgraef}) agree very well with numerical simulations for large values of $d$ (and $A$), capturing the dynamics of the system. When the half-width of the structure $d$ becomes similar to the front width, the ansatz (\ref{eq:kink-antikink}) is not expected to work well. In this case, the numerical simulations indicate that the nullcline $h_+$ is located slightly above the prediction (\ref{walgraef}). 

\begin{figure}
\includegraphics[scale=1.0]{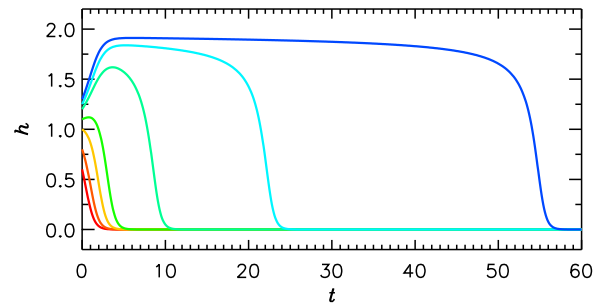}
\caption{(Color online) Time evolution of $h$ for the trajectories plotted in color in Figs.~\ref{PP_d_H} and \ref{PP_A_H} obtained integrating Eq.~(\ref{GL}) starting from an initial condition (\ref{eq:initial_condition}) with $d_0=4.98$ and, from bottom to top, $H_0= 0.6$, $0.8$, $1.0$, $1.1$, $1.2$, $1.25$, $1.3$.} 
\label{time}
\end{figure}

The time evolution of trajectories generated with injected signals of different amplitudes can be observed in Fig.~\ref{time}. Colors correspond to the trajectories shown in in Fig.~\ref{PP_d_H} and  Fig.~\ref{PP_A_H}. For $H_0 \gtrsim 1.1$ trajectories first grow while for $H_0 \lesssim 1.1$ they decay exponentially. In fact, the shape of the trajectories changes gradually and $H\approx 1.1$ is a pseudo-threshold \footnote{Pseudo-thresholds are typical in Type II excitability}.

\begin{figure}[t!]
\includegraphics[width=\columnwidth]{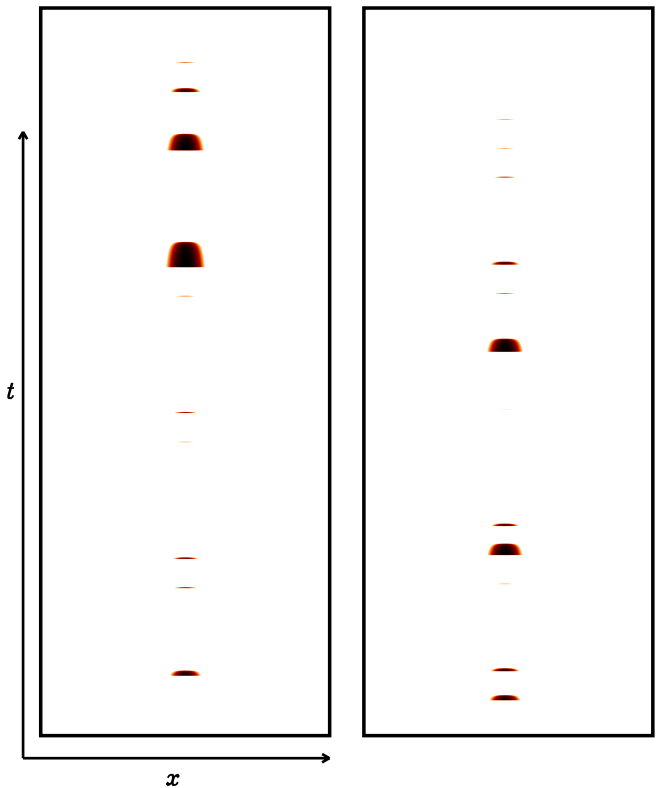}
\caption{(Color online) Dynamics generated by the repetitive addition of Gaussian signals (\ref{prof}) with a noisy amplitude (\ref{eq:noisy_amplitude}) (See text). The range of $x$ plotted is $[-22,22]$ and the range of $t$ is $[0,5000]$ (left panel) and $[5000,10000]$ (right panel).} 
\label{noise}
\end{figure}

Many studies have been devoted to the interplay between excitability and noise \cite{Lindner2004}. The presence of noise  may trigger excitable excursions even for sub-threshold perturbations. We illustrate this effect in our system by injecting signals of the form (\ref{prof}) with a fluctuating amplitude 
\begin{equation}
 G=G^0+\sqrt{D} \xi(t)
 \label{eq:noisy_amplitude}
\end{equation}
where $\xi(t)$ is a Gaussian white noise of zero mean and correlation $<\xi(t) \xi(t')>=\delta(t-t')$. Fig.~5 shows the effect of adding a signal of half-width $\Gamma=15$ and duration $\Delta t=2$ every $T=200$. This allows to plot several events in the same figure. The signals have $G^0=0.79$, below the pseudo-threshold, and are subject to noise with $D=0.3$. Without noise all the perturbations relax fast to $u_-$ and there are no excitable patches. As shown in the figure, in this case the noise plays a constructive role triggering randomly the appearance of patches. We note that for the parameters considered the gap between the steady state and the (pseudo)-threshold is quite large and can not easily be tuned. Therefore, while the results of Fig. 5 show that the noise can have a constructive effect, it is unlikely that, in absence of deterministic perturbations, the noise by itself can trigger excitable excursions, as required, for instance, to observe coherence resonance \cite{PikovskyCohRes}.

To summarize, in the present Letter, we have presented and discussed evidence for a novel mechanism leading to the appearance of transient localized spatiotemporal patches. We have shown that these patches can be understood as excitable excursions. The mechanism presented here only requires the coexistence of two stable homogeneous solutions and spatial coupling such that the fronts connecting the homogeneous solutions are monotonic. We have shown the existence of a pseudo-threshold such that, while sitting on one of the homogeneous states, sub-threshold perturbations decay exponentially. In contrast super-threshold perturbations induce a long excursion. The excursion is characterized by the fast emergence of a structure formed by two back to back fronts connecting the two homogeneous states followed by a slow approximation of the fronts until they eventually annihilate each-other. These two well separated time scales, which do not appear explicitly in Eq.~(\ref{GL}), are an emerging property of the dynamics and allow for a clear observation of the patches.

These patches apparently resemble the excitable localized structures of \cite{Gomila2005}, obtained when a stable oscillatory localized structure disappears through a limit-cycle instability, and the transient localized structures reported for a locally excitable medium \cite{Hecht2010}. From an observational point of view, while transient localized structures normally have a characteristic spatial size independent of the spatial extension of the perturbation, the size of the patches generated here is determined by that of the perturbation. From a fundamental perspective, the mechanism introduced here does not require local excitability nor does it need to support localized structures.

The scenario presented in this Letter is quite different from the two {\it classical\/} types of excitability since it does not require the existence of a nearby oscillatory regime in parameter space. Thus, it could explain experimental observations of transient localized spots in a more general setting. Systems with bistable homogeneous states in which this mechanism could be observed include, for instance, optical systems \cite{Marino2014,*Pesch2007}, chemical reactions \cite{Marts2004,*Marts2006,*Meron2001} or vegetation growth models \cite{Fernandez-Oto2013,*Escaff2015}.

For simplicity, we have focused on the case in which the two homogeneous stable states are equivalent, but the mechanism introduced here can generate excitable patches even if they are not equivalent. Consider the system sitting on the most stable state. A super-threshold perturbation can then similarly trigger a kink-antikink solution connecting the two homogeneous states. If the difference in stability is not too big, the fronts will then approach each-other with a velocity that is small, such that excitable patches are long-lived.

\acknowledgments
This research was supported by the Research Foundation - Flanders (FWO), by the IAP, by the research council VUB and by Ministerio de Econom\'{\i}a y Competitividad (Spain) and Fondo Europeo de Desarrollo Regional under project ESoTECoS FIS2015-63628-C2-1-R (MINECO/FEDER).

\end{document}